\documentclass[aps,floatfix,prl,twocolumn,a4paper,10pt,citeautoscript,
longbibliography,showpacs]{revtex4-1}

\usepackage[english]{babel} 	%%% LaTeX for English
\usepackage{amsmath}        	%%% Mathematics AMS package
\usepackage{amssymb}        	%%% More math symbols
\usepackage{bm}				%%% Bold math font
\usepackage{bbm}            	%%% Font for ensembes

\usepackage{graphicx}       	%%% Include graphics
\usepackage{graphics}       	%%% Include graphics
\usepackage{color}
\usepackage{xcolor}
\DeclareGraphicsExtensions{.png,.jpg,.pdf,.eps} %%% Extensions for pdflatex
\usepackage{hyperref}
\hypersetup{colorlinks=true, pdfstartview=FitV,linkcolor=blue, citecolor=blue, urlcolor=blue}

\definecolor{MS-color}{RGB}{218,39,37}

\newcommand{\bs}{\boldsymbol}
\newcommand{\Eqref}[1]{\eqref{#1}}
\newcommand{\B}{{\bs B}}
\usepackage{xcolor}     %%% Color options for eps

\begin{document}
%%%%%%%%%%%%%%%%%%%%%%%%%%%%%%%%%%%%%%%%%%%%%%%%%%%%%%%%%%%%%%%%%%%%%
%%%%%%%%%%%%%%%%%%%%%%%%%%%%%%%%%%%%%%%%%%%%%%%%%%%%%%%%%%%%%%%%%%%%%
%%%% Title informations and authors
 %\title{Large spontaneous currents in anisotropic $s+is$ superconductors with inhomogeneous pairing}
 %\title{Spontaneous-field polarization sensitive test of the s+is/s+id dichotomy in anisotropic multiband superconductors}
 \title{Polarization of spontaneous magnetic field and magnetic fluctuations 
 in $s+is$ anisotropic multiband superconductors}
 
 \author{V. L. Vadimov}
  \affiliation{Institute for Physics of Microstructures, Russian
 Academy of Sciences, 603950 Nizhny Novgorod, GSP-105, Russia}
\author{M.A. Silaev}
 \affiliation{Department of Physics and Nanoscience Center, University of Jyv\"askyl\"a,
 P.O. Box 35 (YFL), FI-40014 University of Jyv\"askyl\"a, Finland}

\date{\today}

%%%%%%%%%%%%%%%%%%%%%%%%%%%%%%%%%%%%%%%%%%%%%%%%%%%%%%%%%%%%%%%%%%%%%
%%%% The abstract
\begin{abstract}
  We show that multiband superconductors with broken time-reversal symmetry
 can produce spontaneous currents and magnetic fields  in response to the local variations of
 pairing constants.  Considering the iron pnictide superconductor Ba$_{1-x}$K$_x$Fe$_2$As$_2$ as an example 
 we demonstrate that both the point-group symmetric  $s+is$ state and the C$_4$-symmetry breaking $s+id$ states
  produce in general the same magnitudes of spontaneous magnetic fields. 
  In the  $s+is$ state these fields are polarized mainly in ab crystal plane, while in 
  the $s+id$ state their ab-plane and c-axis components are of the same order. 
  The same is true for the random magnetic fields which are produced by the 
   order parameter fluctuations near the critical point of the time-reversal symmetry breaking phase 
  transition. Our findings can be used as a direct test of the $s+is/s+id$ dichotomy and the 
  additional discrete symmetry breaking 
  phase transitions with the help of muon spin relaxation experiments.  
 \end{abstract}

\maketitle

 {\bf Introduction.}
 Superconducting states with spontaneously broken 
 time-reversal symmetry (BTRS)  
 have been recently in the focus of interest. First such states have been studied in connection with the  
chiral p-wave order parameter in the superfluid $^3$He A phase\cite{volovik2009universe} and Sr$_2$RuO$_4$ superconducting compound\cite{Mackenzie.Maeno:03}. 
More recently, $s+id$ and $s+is$ states have been suggested as the candidate order parameters in 
multiband iron pnictide compounds\cite{Lee.Zhang.Wu:09,Zhang2,Thomale,Chubukov2,Johan,SWaveHoleDoped1,SWaveHoleDoped2}. 
Recent experiment\cite{Grinenko2017} supports this hypothesis
demonstrating the presence of spontaneous currents in the ion irradiated 
samples of Ba$_{1-x}$K$_x$Fe$_2$As$_2$ in the certain doping level interval.  

Spontaneous currents were predicted to exist near impurities in $s+id$
superconducting states which spontaneously break the C$_4$ crystalline symmetry of the 
parent compound \cite{Lee.Zhang.Wu:09}. As for the $s+is$ states, initially it has been claimed that 
magnetic field can appear only in samples subjected to strain \cite{ChubukovMaitiSigrist}.
However,  this conclusion was made based on the specific circularly-symmetric model of the impurity.

 More general consideration has shown\cite{Silaev.Garaud.ea:15,Garaud.Silaev.ea:15} that magnetic fields in the $s+is$ state can be 
 generated without strain in the presence of the general-form inhomogeneities of the order parameter.
 They can be induced e.g. by the domain wall between $s+is$ and $s-is$ states\cite{Garaud.Babaev:14}, attached to the sample edge
 or by any external controllable perturbation such as the local heating.
 Later the particular case of two-dimensional defects elongated along the crystal c-axis and forming square shapes in 
 the ab plane have been studied \citep{Lin2016}. In such system the spontaneous magnetic field generated 
 in $s+is$ state is several order of magnitude smaller than in the $s+id$ one.  
  As we show below this difference is not generic and under more general conditions the magnetic field
  amplitudes produced in two states are of the same order.   
  % 
  %%%%%%%%%%%%%%%%%%%%%%%%%%%%%%%%%%%%%%%%%%%%%%%%%%%%%%%%%%%%
 \begin{figure}[!htb]
 \centerline{\includegraphics[width=1.0\linewidth]{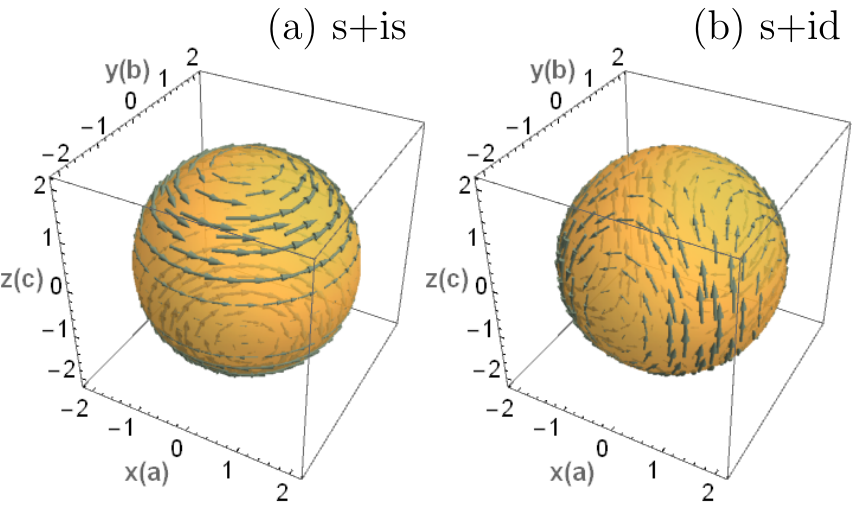}
 \put (-60,100) {$\bm B$}
 \put (-180,100) {$\bm B$} }
 \caption{\label{Fig:Sketch} 
 Arrows show the spontaneous magnetic fields 
generated by rotationally symmetric 3D inhomogeneity 
 of the interband phase difference $\theta_{13}=\theta_{13}(r)$ 
 as given by the Eq.(\ref{Eq:MagneticField}). 
 (a)  Anisotropic $s+is$ state with $\gamma^{x}_{13}= \gamma^{y}_{13}=2\gamma^z_{13}$ 
  and (b) $s+id$ state with $\gamma^z_{13}=0$ and $\gamma^{x}_{13}= - \gamma^{y}_{13}$. 
  The scale of magnetic field is the same in (a,b).
 %Parameters are discussed in text after Eq.(\ref{Eq:mfGL}).
 The field is plotted on the spherical surface $r=const$. The crystal anisotropy axis is $z$.
 The field is rotationally symmetric in $s+is$ state and changes the sign under the $C_4$ rotation in $s+id$ state. }
 \end{figure}
 %%%%%%%%%%%%%%%%%%%%%%%%%%%%%%%%%%%%%%%%%%%%%%%%%%%%%%%%%%%%% 

The purpose of the present paper is threefold. First, we show that the spontaneous magnetic field is generated 
both in the $s+is$ and $s+id$ state due to the general-form inhomogeneities of the pairing interactions. 
Such form of disorder can exist in the sample even without the externally generated defects 
just due to the spatially-inhomogeneous doping level. Second, we demonstrate that in the general case, when the
system is inhomogeneous both in the ab-plane and in the c-direction $s+is$ and $s+id$ states yield the same 
magnitudes of spontaneous fields. 
However, as shown schematically in Fig.(\ref{Fig:Sketch}) this regime is characterized by the 
qualitatively different polarizations of the spontaneous field in $s+id$ and $s+is$ states. 
This prediction can be used for resolving the $s+id/s+is$ dichotomy in real materials with the help
of muon spin relaxation experiments\cite{Sonier2000,PhysRevB.89.020502,Grinenko2017}. 
 Third, we demonstrate that the order parameter fluctuations
near the BTRS phase transition generate random magnetic fields with the critical correlation radius.
Thus, the discrete symmetry breaking phase transition  can be revealed through the 
magnetic field fluctuations.
%%%%%%%%%%%%%%%%%%%%%%%
     
 {\bf Three-band model.} Here we develop general treatment of spontaneous magnetic fields in
  BTRS states further considering inhomogeneities created by the spatial variation of pairing 
  constants in the minimal three-band microscopic model 
 \cite{StanevTesanovic,Benfatto,Chubukov2} with three distinct superconducting 
gaps $\Delta_{1,2,3}$ residing in different bands. The pairing which leads to the BTRS 
state is dominated by the competition of two interband repulsion channels $\eta_{1,2}>0$
described by the following coupling matrix  %$\hat g = - \nu_0 \hat\Lambda$, where
  \begin{equation}
 \label{Eq:Model3BandText}
 \hat\Lambda =  - \nu_0 \left(%
 \begin{array}{ccc}
 0 & \eta_1 & \eta_2   \\
 \eta_1 & 0 & \eta_2   \\
 \eta_2 & \eta_2 & 0         \\
 \end{array}  \right) \,.
 \end{equation}
We assume for simplicity that the density of states $\nu_0$ is the same 
in all superconducting bands. 
 This model can be used for both the  $s+is$ and $s+id$ states. In the former case $\Delta_{1,2}$ 
correspond to the gaps at hole pockets and $\Delta_3$ is the gap at electron 
pockets, so that $u_{hh}= \nu_0\eta_1$ and $u_{eh} = \nu_0\eta_2$ are respectively 
the hole-hole and electron-hole interactions \cite{Benfatto,Chubukov2}. The same model 
\Eqref{Eq:Model3BandText} can be used to describe the $s+id$ states but there, 
$\Delta_{1}$ and $\Delta_{2}$ describe gaps in $(0,\pm\pi)$ and $(\pm\pi,0)$ electron pockets, respectively.
In the hole pocket the gap is $\Delta_3$, so that $u_{eh} = \nu_0\eta_1$ and $u_{ee} = \nu_0\eta_2$ are electron-hole
and electron-electron interactions respectively \citep{Garaud2017,Garaud.Silaev.ea:15}. 

The inhomogeneities of pairing interactions in the model (\ref{Eq:Model3BandText}) produce spatially varying gap amplitudes 
$|\Delta_i|$ and phases $\theta_i$.
 Their gradients can generate spontaneous magnetic field 
 according to the modified London expression in multiband superconductors\cite{Silaev.Garaud.ea:15}
 \begin{equation} \label{Eq:MagneticField}
 \B = - 4\pi \nabla\times \left( \hat\lambda^{2}_L {\bm j} \right) +
 \frac{1}{\tilde e N}\sum_{k>i} \nabla\times\left( \hat\gamma_{ki} \nabla\theta_{ki}\right),
 \end{equation}
 where we use the units with $\hbar = c=1$ and introduce interband phases
  $\theta_{ki}=\theta_{k}-\theta_i$. % flux quantum is $\Phi_0 = 2\pi/e$
 Here the London penetration depth is given by 
 $  \hat \lambda^{-2}_L = \sum_k \hat \lambda^{-2}_{k}$ and
 $  \hat\gamma_{ki}= \hat\lambda_L^2 ( \hat\lambda_k^{-2} - \hat \lambda_i^{-2}) $
 , where  
 $\hat\lambda_{k}$ are in general the tensor coefficients characterizing 
 contribution of each band to the Meissner screening. In the clean limit they can be expressed 
 as follows 
 \begin{equation} \label{Eq:LondonK}
 \hat\lambda^{-2}_{k} = 8 \pi \rho {\tilde e}^2 \hat K_k |\Delta_k|^2 ,
 \end{equation}
 where  $\hat K_{k} = \langle \bm v_{k} \bm v_{k} \rangle$ is the anisotropy tensor, 
  $\bm v_{k} $  is the Fermi velocity in $k$-th band normalized to the certain band-independent 
  characteristic velocity $\bar v_F$. 
   We normalize the gaps by $T_c/\sqrt{\rho}$, where $\rho =\sum_n \pi T_c^3\omega_n^{-3} \approx 0.1$ and 
  $T_c$ is the critical temperature, magnetic field by $B_0 = T_c\sqrt{\nu_0/\rho}$, which is close to 
  the thermodynamic critical field at zero temperature\cite{Saint-James.Thomas.ea}. 
  Length is normalized by the Cooper pair size $\xi_0 =\bar v_F/T_c$, and we introduce the dimensionless Cooper pair charge 
  is $\tilde e =  2 e \xi_0^2 B_0 $. 

  In contrast to the usual London electrodynamics, the multicomponent superconducting systems 
 can generate spontaneous magnetic fields due the second term in Eq.(\ref{Eq:MagneticField}) acting 
 as a source according to the mechanisms described below.  
  The source term in Eq.(\ref{Eq:MagneticField}) is non-zero if $\nabla\theta_{ki}\neq 0$
    and tensors $\hat\gamma_{ki}$ are constant in space but anisotropic.
     This scenario is generic for the $s+id$ state when all three  components $\gamma^{x,y,z}_{ki}$
     are different. 
     {
  The $s+is$ state is isotropic in ab plane, but there is anisotropy in {ca and cb }
  planes  $\gamma_{ki}^x = \gamma_{ki}^y \neq \gamma_{ki}^z$. In this case the ab-plane inhomogeneities 
  are decoupled from the magnetic field \cite{Lin2016}, so that $B_z=0$. 
  However, in general the systems is 
  inhomogeneous along the c-axis direction as well, 
  which yields the magnetic response $B_{x,y}$ of the same magnitude as the $s+id$ state. 
  
  The general field structures produced by the 3D inhomogeneity in $s+is/s+id$ states
 can be found using Eq.(\ref{Eq:MagneticField}). In Fig.\ref{Fig:Sketch}  we show spontaneous field produced by 
  the interband phase difference modulation $\theta_{13}(r)$. 
  On the $s+is$ case only $B_{x,y}\neq 0$  while in
   $s+id$ state the field has all components. 
   }   
 
 {\it Second}, the component $B_z\neq 0$ can be generated even in $s+is$ state\cite{Garaud.Babaev:14,Garaud.Silaev.ea:15,Silaev.Garaud.ea:15,Lin2016}.  
 According to Eq.(\ref{Eq:MagneticField}) for that we need
  simultaneously $\nabla_{x,y}\theta_{ki}\neq 0$ and $\nabla_{x,y}\gamma^{x,y}_{ki}\neq 0$ with additional requirement that these
  gradients are non-collinear to each other.
 Therefore $B_z$ component in $s+is$ state 
 is significantly smaller than in the $s+id$ state, where only $\nabla_{x,y}\theta_{ki}\neq 0$ is needed. 
 Therefore  the largest 
   spontaneous  field in the $s+is$ case appears in the direction perpendicular to the anisotropy axis, while in
   $s+id$ all components are of the same order. On can distinguish between these states by analysing the polarization of spontaneous magnetic fields with the help of the muon spin relaxation techniques\cite{Sonier2000,PhysRevB.89.020502,Grinenko2017}. 
     Below we illustrate these conclusions using more detailed calculations close to $T_c$ using the Ginzburg-Landau (GL) theory.  
 
 {\bf Ginzburg-Landau calculation.} 
 To go beyond local approximation we can calculate spontaneous magnetic fields using GL theory derived for $s+is/s+id$ states \cite{Garaud2017,Garaud.Silaev.ea:15} corresponding 
 to the model (\ref{Eq:Model3BandText}). 
 The general free energy density, normalized to $B_0^2$ is given by  $F= F_s + B^2/8\pi$, where 
 the GL free energy describing both the $s+is$ and $s+id$ states is given by
 \begin{align}
 & F_s = \sum_{j=1}^2 
 \left(
  ( \bm{\hat\Pi} \psi_j )^* \hat k_{jj} (\bm{\hat\Pi} \psi_j) + \alpha_j|\psi_j|^2 + \frac{\beta_j}{2}|\psi_j|^4
 \right)
 \nonumber \\
 & + 2 ( \bm{\hat\Pi} \psi_1)^* \hat k_{12} \bm{\hat\Pi} \psi_2 
 + \gamma|\psi_1|^2|\psi_2|^2 + \delta \psi_1^{*2}\psi_2^2 + c.c ,
 \label{Eq:FreeEnergy}
 \end{align}
 where $\bm{\hat\Pi} = \nabla - i\tilde{e}\bm A$.
 This model is formulated in terms of the two order parameters  $\psi_1$ and $\psi_2$ which are 
 related to the individual gap functions within separate bands as  
 $(\Delta_1,\Delta_2,\Delta_3)=(\zeta\psi_2-\psi_1,\zeta\psi_2+\psi_1,\psi_2)$, where
 $\zeta=(\eta_1-\sqrt{\eta^2_1+8\eta^2_2})/4\eta_2$. 

 The coefficients of gradient terms in Eq.(\ref{Eq:FreeEnergy}) are combined 
 from the anisotropy tensors characterising each superconducting band as follows
   {
 \begin{align}\label{Eq:k11}
  & \hat k_{11} = \rho(\hat K_1 + \hat K_2)  
   \\ \label{Eq:k22}
  & \hat k_{22} = \rho[\zeta^2(\hat K_1 + \hat K_2) + \hat K_3] 
   \\ \label{Eq:k12}
  & \hat k_{12} = \zeta\rho( \hat K_2 - \hat K_1 ) 
  \end{align}
  }
 The difference between $s+is$ and $s+id$ symmetries is determined by the 
structure of mixed-gradient coefficients (\ref{Eq:k12}) in ab plane. 
That is for $s+is$ state $K_i^{x}=K_i^{y}\equiv K_i^{xy}$ so that 
$k_{12}^x=k_{12}^y\equiv k_{12}^{xy}$. For
$s+id$ state $K_{1,2}^{x}\neq K_{1,2}^{b}$ but $K_1^{x}= K_2^{y}$ so that
 $k_{12}^x= - k_{12}^y\equiv k_{12}^{xy}$. 
Despite having quite different properties in the ab plane both states are characterized by the 
  anisotropy in ca and cb planes determined by the coefficients $  K_i^z \neq K_i^{x}, K_i^{y}$. 
  In $s+is$ state this anisotropy provides linear coupling between
  magnetic field and pairing constant inhomogeneities. For that 
  at least two bands should  have different anisotropies,
 otherwise the problem can be rescaled to the fully isotropic
 one when only the non-linear coupling is possible yielding much smaller spontaneous currents. 
 %
  
 %%%%%%%%%%%%%%%%%%%%%%%%%%%%%%%%%%%%%%%%%%%%%%%%%%%%%%%
 \begin{figure}[htb!]
 \centerline{$
 \begin{array}{c}
 \includegraphics[width=1.0\linewidth]{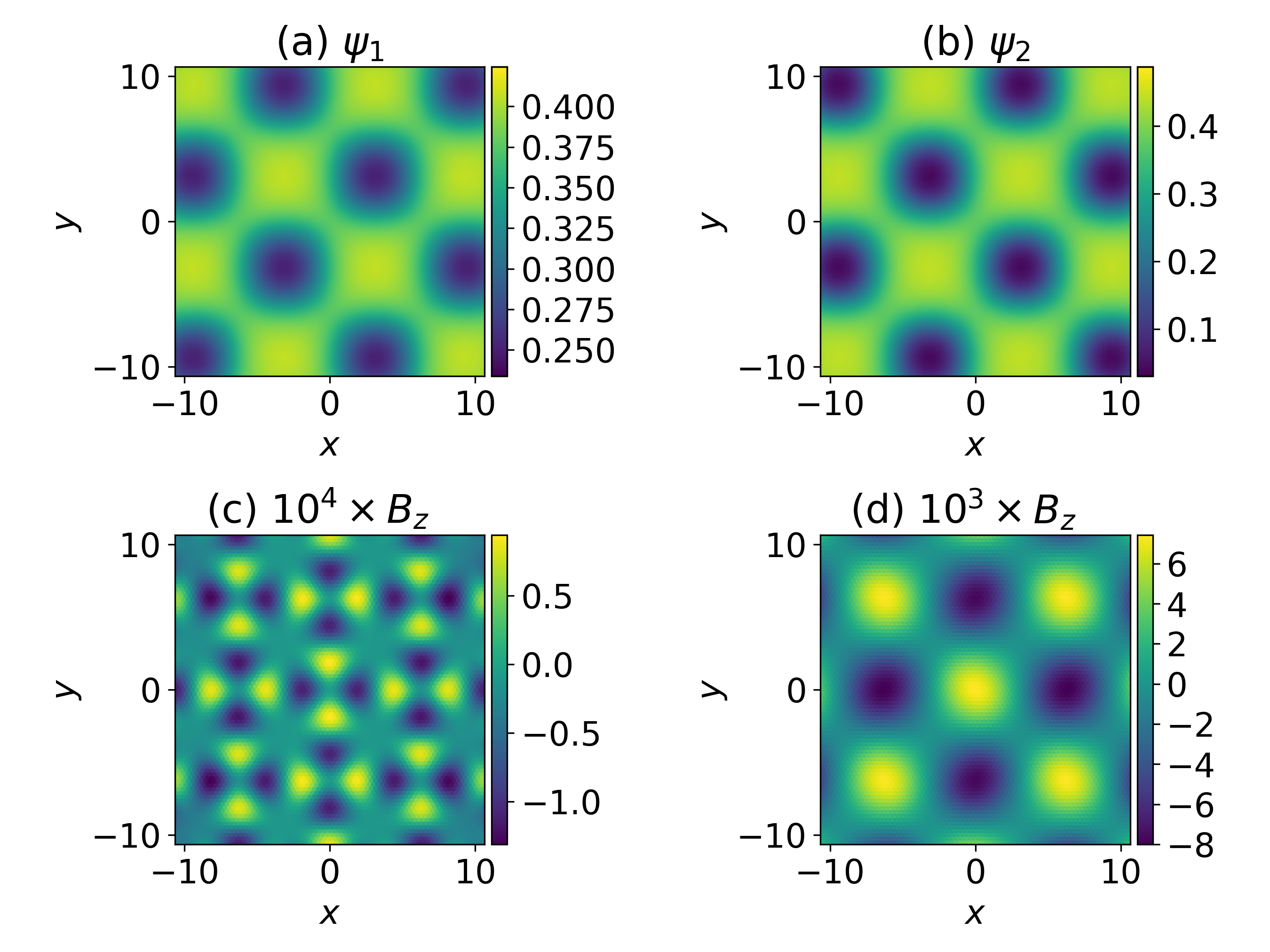}
 \end{array}$}
 \caption{\label{Fig:2} (Color online) 
  (a,b) Order parameter modulation produced by the ab-plane inhomogeneities of the form (\ref{Eq:Lattice}). 
  The corresponding spontaneous field $B_z$ is shown for (c)  $s+is$ state
  with  $K^{xy}_1=1$, $K^{xy}_2=1.5$, $K^{xy}_3=0.5$ and (d) $s+id$ state with
  $K^x_1 = K^y_2 = 1$, $K^y_1=K^x_2=1.5$ and $K^{xy}_3=0.5$.
  { GL parameter is $\tilde{e}=1/4$} , $\tau = 0.2$, the field is
normalized to $\tau B_0 / \tilde e $. 
   }
 \end{figure}   
% %%%%%%%%%%%%%%%%%%%%%%%%%%%%%%%%%%%%%%%%%%%%%%%%%%%%%

% %%%%%%%%%%%%%%%%%%%%%%%%%%%%%%%%%%%%%%%%%%%%%%%%%%%% 
 \begin{figure}[htb!]
\includegraphics[width=\linewidth]{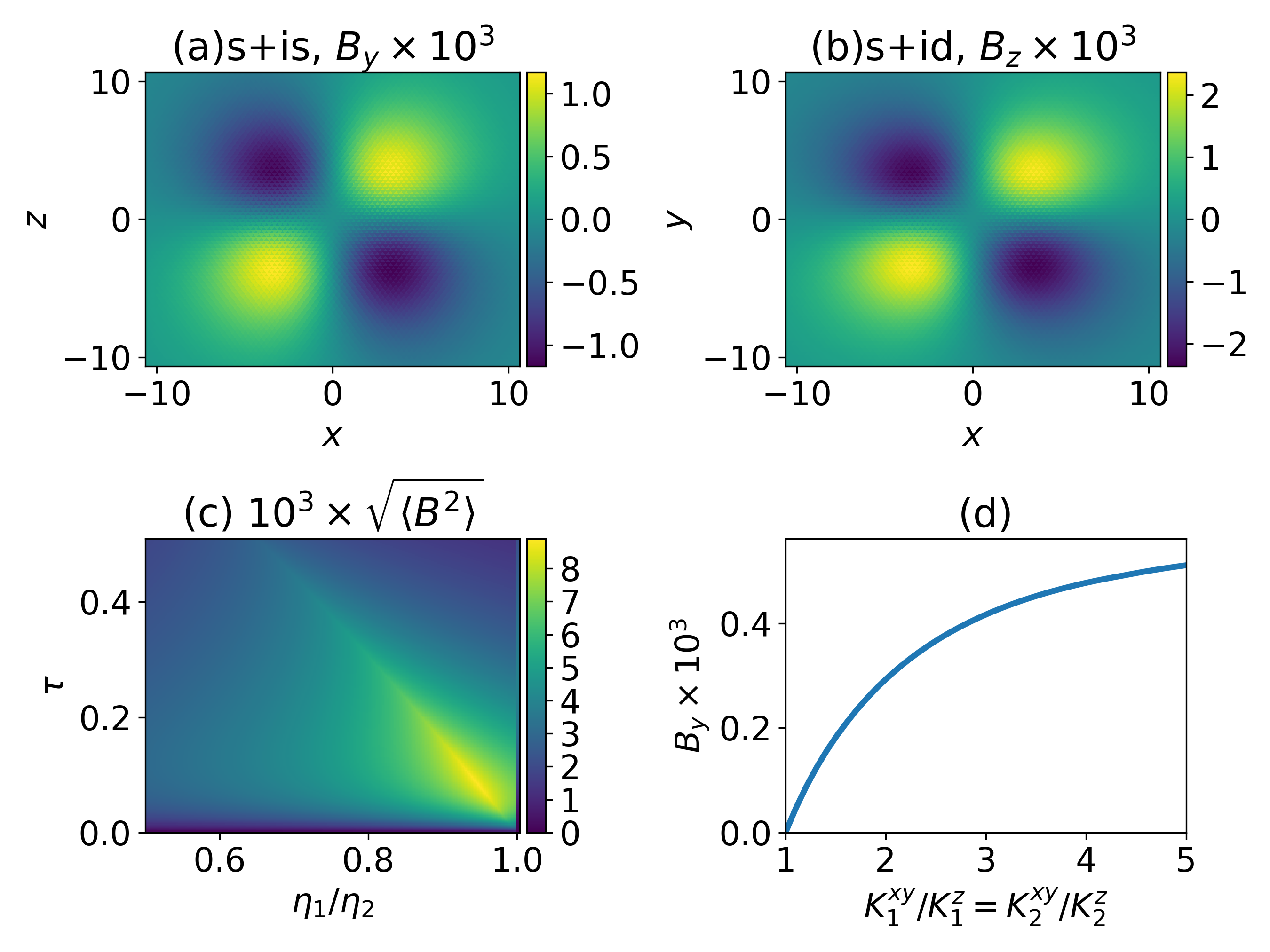}
 \caption{\label{Fig:3} 
 (a,b) Spontaneous fields produced by the inhomogeneities of the 
 pairing constant $\eta_2(\bm r)$ (\ref{Eq:gaussianXY},\ref{Eq:gaussianXZ}) and $\eta_1 =1$. 
(a) ca-plane defect in $s+is$ superconductor with anisotropy parameters for $s+is$ are 
 $K^{xy}_1=1$, $K^{xy}_2=1.5$, $K^{xy}_3=0.5$, $K^z_{1} = 2$, $K^z_{2}=3$,
 $K^z_{3}=0.5$.  
{(b) ab-plane defects in $s+id$ state characterized by
$K^{x}_1=K_2^y=1$, $K^{y}_1=K^x_2=1.5$, $K^{x}_3=K^{y}_3=0.5$}.
GL parameter {$\tilde{e}=1/4$}
  for both cases, $\tau = 0.2$. The field is
normalized to $\tau B_0 / \tilde e $. 
(c) Spontaneous magnetic field in the ab-plane  produced by the critical fluctuations near the BTRS transitions 
in anisotropic $s+is$ state. 
The field is measured in $ (|k_{12}^z - k_{12}^{xy}|/\sqrt{k_{11}k_{22}}) (T_c/E_F)B_0$, where 
$E_F$ is the Fermi energy. The field maximum at each value of 
$\eta_1/\eta_2$  corresponds to the BTRS critical temperature.
(d) The magnitude of the in-plane magnetic field vs. the effective mass
anisotropy in the $s+is$ superconductors.
 }
 \end{figure}    
% %%%%%%%%%%%%%%%%%%%%%%%%%%%%%%%%%%%%%%%%%%%%%%%%%%%%%%%%5
 
 The other coefficients in GL expansion (\ref{Eq:FreeEnergy}) are 
 expressed in terms of the pairing constants (\ref{Eq:Model3BandText}) as 
 \begin{align}
 &\alpha_{1} = -2(G_0-G_{1}+\tau )
 \\
 &\alpha_{2} = -( 1+2\zeta^2)(G_0-G_{2}+\tau ) 
 \\
 &\beta_{1} = 2 ;\; \beta_{2}=1+2\zeta^4 
 \\
 &\gamma = 4\zeta^2 ; \; \delta=2\zeta^2
 \end{align}
 where $\tau=1-T/T_c$, $G_1 = 1/\eta_1$ and $G_2 =\left(\eta_1 + \sqrt{\eta^2_1 + 8\eta^2_2} \right) /4\eta_2^2$
are the positive eigenvalues of the matrix (\ref{Eq:Model3BandText})
 $ \hat\Lambda^{-1}$ and $G_0 = \min(G_1,G_2)$.
 {The Ginzburg-Landau model is valid in the vicinity of $T_c$ when both order parameters $|\psi_{1}|$
 and $|\psi_{2}|$ are small. In this regime the bulk BTRS state appears for the close values of pairing 
 constants due to the following reason. 
 In the homogeneous state  both order parameters appear simultaneously when $\eta_1=\eta_2$.
 In case of the finite detuning, one of the order parameters nucleates first. E.g. when $\eta_1> \eta_2$ the $\psi_1$ state nucleates at $T_c$ since  $G_0=G_1<G_2$.  Then the bulk critical temperature of the BTRS transition, that is nucleation 
 of $\psi_2$ in this case  can be found from the relation $\alpha_2 = -2\zeta^2 |\psi_1|^2$, which is equivalent to 
 $\tau = G_2 - G_1 + \zeta^2 |\psi_1|^2$. The we have the restriction $|G_1-G_2| \leq \tau$ so that $|1-\eta_1/\eta_2| \leq \tau$. 
 The inhomogeneous BTRS state however  occurs for much larger amplitudes of the pairing constant variations because the 
 additional order parameter nucleates locally at the regions where $\eta_1\approx \eta_2$.      }

 Below, we consider the spontaneous magnetic field produced by the superconducting currents generated by the 
 inhomogeneities of the pairing constant $\eta_2=\eta_2(\bm r)$. 
 %%%%%
  At first let us consider the 2D inhomogeneities in the ab plane, so that $\eta_1=1$ and 
 \begin{equation}\label{Eq:Lattice}
 \eta_2(x,y) = 1 + 0.5\sin(x/2)\sin(y/2).
 \end{equation} 
 This model allows for demonstrating differences between the linear and non-linear mechanisms of the 
 spontaneous  current generation, where the former takes place for $s+id$ and the latter is $s+is$ pairings.
 The magnetic field produced by ab-inhomogeneities has only z-component.
 The calculated distributionc of $B_z=B_z(x,y)$ is shown in Figc.\ref{Fig:2}c,d.
Where  
 the magnetic field is given in the units of $\tau B_0/\tilde e$ which has an order of
 the upper critical field $H_{c2}$ at a given temperature. One can see that for one and the same set of parameters 
 $s+is$ state yields the spontaneous magnetic field response about $10^2$ times smaller than $s+id$, which is consistent    
 with the results obtained before \cite{Lin2016}. 
 
 Except of the special case of ab-plane inhomogeneities, in general the $s+is$ and $s+id$ states
 produce the magnetic fields of comparable amplitudes. To demonstrate this we compare responses produced by the 
  pairing constant variation given by
 %%%%
 \begin{align} \label{Eq:gaussianXY}
 \eta_2 =1 + 0.5 e^{-(x^2+y^2)/8} & & \textnormal{ $s+id$ state   } 
 \\
 \label{Eq:gaussianXZ}
 \eta_2 = 1+0.5 e^{-(x^2+z^2)/8} & & \textnormal{ $s+is$ state } 
 \end{align}
 %%%%%%%%%%%
 The former inhomogeneity (\ref{Eq:gaussianXY}) corresponds to the ab-plane defect, while the latter (\ref{Eq:gaussianXZ})
 is the ca-plane defect.  
 To obtain spontaneous fields produced by ca-plane  defects in $s+is$ case we assume that there is 
 {ca-plane anisotropy} set by the choice of coefficient ratio in different bands
 $K^{xy}_1=1$, $K^{xy}_2=1.5$, $K^{xy}_3=0.5$, $K^z_{1} = 2$, $K^z_{2}=3$,
 $K^z_{3}=1$. Such system yields the magnetic field component $B_y(x,z)$ shown in Fig.\ref{Fig:3}a.
   One can compare it with the qualitatively similar distribution of $B_z$ component produced by the Gaussian 
   ab-plane inhomogeneity 
   (\ref{Eq:gaussianXY}) in the $s+id$ state shown in Fig.\ref{Fig:3}b. 
 {Magnetic signatures of ca-defect in $s+id$ states are  qualitatively similar to that in  $s+is$ shown in 
 Fig.\ref{Fig:3}a.
 In ca-plane the $s+id$ state  is described by structurally identical GL equations as the $s+is$ one 
 with the interchange  $K_k^{xy}\to K_k^x $.  
 The 122 iron pnictide compounds has been shown to feature anisotropy which  can vary from in rather wide limits 
 from $ K_i^{xy} / K_i^{z} \approx 1-3$  \citep{Yuan2009} to $ K_i^{xy} / K_i^{z} \approx 4-5$ \cite{PhysRevB.89.134502}. 
 In Fig.\ref{Fig:3}d we show the field amplitude dependence on the degree of anisotropy
  $K_1^z / K_1^{xy} = K_2^z / K_2^{xy} \neq 1$ and fixing $K_3^z / K_3^{xy} =1$.   }   
   
  {
   The general analytical expression for spontaneous field in case of the 3D inhomogeneity can be obtained 
   using the reduced GL theory in the vicinity of BTRS transition. It can be constructed }
   assuming the main order parameter to be $\psi_1 =|\psi_1|e^{i\varphi_1}$ with 
    $|\psi_1|=const$ and introducing the 
   BTRS order parameter 
    $\Theta = -i (\psi_2\psi_1^*)/|\psi_1| $. 
   Then we represent the GL free energy in terms if the  
   gauge-invariant momentum $\bm Q = \bm A-\nabla \varphi_1/\tilde{e}$, the real and imaginary parts of the complex order
    parameter  $\Theta= \Theta_r + i \Theta_{im}$ as follows
    \begin{align} \nonumber
    & F(\Theta, \bm Q) = \tilde \alpha_r \Theta^2_r  + \tilde \alpha_{im}\Theta^2_{im} + 
    \frac{\beta_2}{2}|\Theta|^4 + \frac{|\nabla \times \bm Q|^2}{8\pi} + 
    \\ \nonumber    
    & |\psi_1|^2 \tilde e^2 \bm Q \hat k_{11} \bm Q + 2 |\psi_1| \tilde e \bm Q \hat k_{12} 
    \left(\nabla \Theta_r + \tilde e \bm Q \Theta_{im}\right)+ 
    \\ 
    & \left( \nabla + i\tilde e \bm Q \right) \Theta^\ast \hat
    k_{22} \left(\nabla - i\tilde e \bm Q\right) \Theta .
    \end{align}   
   Here $\tilde \alpha_r = \alpha_2 + |\psi_1|^2(\delta - \gamma)$ and 
   $\tilde \alpha_{im} = \alpha_2 + |\psi_1|^2(\delta + \gamma)$. Equation $\tilde
   \alpha_r(T) = 0$ gives the critical temperature of BTRS transition. In the vicinity
   of this transition only the variation of
   $\Theta_r$ is important as the $\tilde \alpha_{im}$ is positive and non-vanishing.
   Therefore we can describe the time-reversal symmetry breaking phase transition 
   in terms of the real-valued order parameter $\Theta_r$:
   \begin{align}  \nonumber
   &
   F(\Theta_r,\bm Q) = \tilde e^2 |\psi_1|^2  \bm Q \hat k_{11} \bm Q + 
   \tilde e^2\Theta_r^2 \bm Q \hat k_{22}\bm Q +  \frac{|\nabla \times \bm Q|^2}{8\pi} 
   \\ 
   \label{Eq:FGLZ2}
   & 
   \tilde \alpha_r \Theta_r^2 + \frac{\beta_2}{2}\Theta_r^4 + 
   \nabla \Theta_r \hat k_{22} \nabla \Theta_r + 2 \tilde e|\psi_1|  \bm Q \hat k_{12} \nabla \Theta_r .  
   \end{align}   
   Note that the real order parameter $\Theta_r$ is still coupled to the magnetic field because 
   the superconducting current obtained from functional (\ref{Eq:FGLZ2}) is given by 
   {
   \begin{equation} \label{Eq:current}
   \bm j = - 2|\psi_{1}|^2 \tilde{e}^2  \hat k_{11} \bm Q-%2 \eta_r^2\tilde e^2 \hat k_{22} \bm Q -
      2\tilde{e}|\psi_{1}| \hat k_{12} \nabla\Theta_r.
   \end{equation}
    }   
   For simplicity let us assume that the coefficients $\hat k_{ii}$ for $i=1,2$ are isotropic and the anisotropy 
   is determined by 
   $\hat k_{12}$. Then, going to the Fourier transform $\Theta_r (\bm r) = V \int e^{i\bm q\bm r} 
   \Theta_r (\bm q) d^3\bm q/(2\pi)^3$ in the 
    volume  $V$  we obtain the magnetic field
   \begin{equation}\label{Eq:mfGL}
   \bm B({\bm q}) =  \Theta_r({\bm q})\sqrt{8\pi/k_{11}}
   (\bm q\times \hat k_{12} \bm q ) /[\lambda (q^2+\lambda^{-2})] ,
   \end{equation}      
    where $\lambda =1/(\sqrt{8\pi k_{11}}|\psi_{1}| \tilde{e}) $ is the London penetration length.
   Eq.(\ref{Eq:mfGL}) shows that gradients $\Theta_r$ with necessity produce the 
   spontaneous magnetic field. 
   They can induced by the inhomogeneous pairing constant through the spatially varying  
   coefficient $\alpha_r=\alpha_r(\bm r)$ in Eq.(\ref{Eq:FGLZ2}). 
   One can see that in the wide range of parameters $q\lambda\gg 1$ magnetic field amplitude is independent on the
   inhomogeneity scale.    
   
   The fields produced by rotationally symmetric 3D defect
   $\Theta_r = \Theta_r(r)$ have the same structure as shown in Fig.(\ref{Fig:Sketch}). 
         Based on the above analysis one can suggest the polarization-sensitive test of the superconducting state symmetry based.
   That is, under general conditions, the spontaneous magnetic field in $s+is$ state is  directed mostly in the ab-plane, 
   with the typical ratio of components $B_z/B_{\perp} \sim 10^{-2}$ as one can see comparing 
   Figs.\ref{Fig:2}c and \ref{Fig:3}a, where  $\bm B_\perp = (B_x, B_y, 0)$.
    On the other hand, $s+id$ state produces spontaneous fields which have  in general all 
   components with the same order $B_z/B_{\perp} \sim 1$.  
   
   {\bf Critical magnetic fluctuations.} 
   Spontaneous magnetic field produced by the order parameter inhomogeneities allows for the direct observation 
   of the critical phenomena and fluctuations near the BTRS phase transition.    
    {   
   From (\ref{Eq:mfGL}) we  get the 
   variance of magnetic field components in ab-plane 
   \begin{align} \label{Eq:DispersionBy}
    \langle B_\perp^2({\bm q}) \rangle = 
   \langle \Theta_r^2({\bm q}) \rangle \frac{(k_{12}^z - k_{12}^{xy})^2}{k_{11}} 
   \frac{8\pi\lambda^2 q_\perp^2q_z^2}{(\lambda^2 q^2 + 1)^2},
   \end{align}
   where $q_\perp = \sqrt{q_x^2 + q_y^2}$. }      
   For simplicity we consider the limiting case  when the cross-coupling gradient terms in the functional 
   (\ref{Eq:FGLZ2}) are rather small $ k_{12}^z,k_{12}^{xy} \ll \sqrt{k_{11}k_{22}} $ when the feedback of 
   magnetic field fluctuations
   can be neglected. Then, fluctuations of the order parameter $\Theta_r$ near the BTRS critical temperature 
   can be calculated using the conventional expression \cite{landau2013statistical}
   $\langle \Theta_r^2({\bm q}) \rangle = T/[2B_0^2 V(k_{22} q^2 + |\tilde\alpha_r| )]$ .
  { Now, we can calculate the average value of the  spontaneous magnetic field amplitude
   $\langle  B_\perp^2 \rangle = V\int d^3\bm q \langle B_\perp^2({\bm q}) \rangle $ 
   using the ultraviolet cutoff at the scale $\xi_0^{-1}$. The dependence of the average 
   amplitude $\bar B_\perp = \sqrt{\langle  B_\perp^2 \rangle}$ on system parameters $(T, \eta_1/\eta_2)$
   for fixed $\eta_1=0.5$ is shown in Fig.\ref{Fig:3}c.
   Using the typical value of $T_c/E_F =10^{-3}$ one can see that the field amplitude in Fig.\ref{Fig:3}c
   is about $10^{-5} B_0$ which is of the same order as produced by $s+is$ state with 
   ab-inhomogeneities shown in Fig.\ref{Fig:2}c.

   The average amplitudes of magnetic field components $\bar B_k^2 = \int B_k^2 n(B_k) dB_k$ can be derived from the
   magnetic field distribution function $n(B_k)$ which is a directly measurable experimental quantity. 
   It can be obtained as the Fourier transform of the complex 
   muon spin polarization function in time domain \cite{Sonier2000}. In this way, comparing the 
   signals from muon beams polarized alon c axis and in the ab plane one can determine whether the system is in  
   $s+is$ or in $s+id$ state. Besides that one can distinguish the line of the  BTRS phase transition. 
   As shown in  Figs.\ref{Fig:3}c the BTRS phase transitions correspond to the distinct maxima of 
   the fluctuating field amplitude. 
   }
   These spontaneous fields provide therefore the direct access to the previously hidden 
   critical behaviour near the discrete symmetry-breaking phase transitions. 
   %%%%%%%%%%%%%%%%%%%%%%%%%%%%%%%%%%%%%%%%%%%%%%%%%%%%%%%%%%%%%%%%%%%%%
   
   {\bf Conclusion.}
      To summarize, we have shown that in general  the $s+id$ and $s+is$ phases in multiband superconductors 
      can produce spontaneous currents and magnetic fields in response to the spatial inhomogeneities caused
      by either the fluctuations of the pairing constants or the critical fluctuations of the order parameter components. 
      This is in contrast to the previous predictions that $s+is$ state has much weaker magnetic signatures. 
      However, the spontaneous field polarization is found to be drastically different in $s+is$ and $s+id$ states 
      making it possible to 
      distinguish between them experimentally using muon spin relaxation measurements. 
      The random magnetic fields produced by the 
      scalar order parameter fluctuations can reveal the critical behaviour near the 
       BTRS transition and in general any additional discrete-symmetry breaking phase transition deep in 
       the superconducting state.   

    {\bf Acknowledgements.}
      We thank Vadim Grinenko, Egor Babaev, Julien Garaud and Alexander Mel'nikov 
      for illuminating discussions. This work was supported by the Academy of Finland (Project No. 297439), 
      Russian Science Foundation (Grant No. 17-12-01383), Russian Foundation
      for Basic Research (Grants no. 17-52-12044 and 18-02-00390) and Foundation for the advancement of theoretical physics
      ``BASIS'' No. 109.  

%%%%%%%%%%%%%%%%%%%%%%%%%%%%%%%%%%%%%%%%%%%%%%%%%%%%%%%%%%%%%%%%%%%%%
%%%% Bibliography
\bibliographystyle{apsrev4-1}
\bibliography{Thermophase}

\begin{thebibliography}{24}%
\makeatletter
\providecommand \@ifxundefined [1]{%
 \@ifx{#1\undefined}
}%
\providecommand \@ifnum [1]{%
 \ifnum #1\expandafter \@firstoftwo
 \else \expandafter \@secondoftwo
 \fi
}%
\providecommand \@ifx [1]{%
 \ifx #1\expandafter \@firstoftwo
 \else \expandafter \@secondoftwo
 \fi
}%
\providecommand \natexlab [1]{#1}%
\providecommand \enquote  [1]{``#1''}%
\providecommand \bibnamefont  [1]{#1}%
\providecommand \bibfnamefont [1]{#1}%
\providecommand \citenamefont [1]{#1}%
\providecommand \href@noop [0]{\@secondoftwo}%
\providecommand \href [0]{\begingroup \@sanitize@url \@href}%
\providecommand \@href[1]{\@@startlink{#1}\@@href}%
\providecommand \@@href[1]{\endgroup#1\@@endlink}%
\providecommand \@sanitize@url [0]{\catcode `\\12\catcode `\$12\catcode
  `\&12\catcode `\#12\catcode `\^12\catcode `\_12\catcode `\%12\relax}%
\providecommand \@@startlink[1]{}%
\providecommand \@@endlink[0]{}%
\providecommand \url  [0]{\begingroup\@sanitize@url \@url }%
\providecommand \@url [1]{\endgroup\@href {#1}{\urlprefix }}%
\providecommand \urlprefix  [0]{URL }%
\providecommand \Eprint [0]{\href }%
\providecommand \doibase [0]{http://dx.doi.org/}%
\providecommand \selectlanguage [0]{\@gobble}%
\providecommand \bibinfo  [0]{\@secondoftwo}%
\providecommand \bibfield  [0]{\@secondoftwo}%
\providecommand \translation [1]{[#1]}%
\providecommand \BibitemOpen [0]{}%
\providecommand \bibitemStop [0]{}%
\providecommand \bibitemNoStop [0]{.\EOS\space}%
\providecommand \EOS [0]{\spacefactor3000\relax}%
\providecommand \BibitemShut  [1]{\csname bibitem#1\endcsname}%
\let\auto@bib@innerbib\@empty
%</preamble>
\bibitem [{\citenamefont {Volovik}(2009)}]{volovik2009universe}%
  \BibitemOpen
  \bibfield  {author} {\bibinfo {author} {\bibfnamefont {G.}~\bibnamefont
  {Volovik}},\ }\href {https://books.google.fi/books?id=6uj76kFJOHEC} {\emph
  {\bibinfo {title} {The Universe in a Helium Droplet}}},\ International Series
  of Monographs on Physics\ (\bibinfo  {publisher} {OUP Oxford},\ \bibinfo
  {year} {2009})\BibitemShut {NoStop}%
\bibitem [{\citenamefont {Mackenzie}\ and\ \citenamefont
  {Maeno}(2003)}]{Mackenzie.Maeno:03}%
  \BibitemOpen
  \bibfield  {author} {\bibinfo {author} {\bibfnamefont {A.~P.}\ \bibnamefont
  {Mackenzie}}\ and\ \bibinfo {author} {\bibfnamefont {Y.}~\bibnamefont
  {Maeno}},\ }\href {\doibase 10.1103/RevModPhys.75.657} {\bibfield  {journal}
  {\bibinfo  {journal} {Rev. Mod. Phys.}\ }\textbf {\bibinfo {volume} {75}},\
  \bibinfo {pages} {657} (\bibinfo {year} {2003})}\BibitemShut {NoStop}%
\bibitem [{\citenamefont {Lee}\ \emph {et~al.}(2009)\citenamefont {Lee},
  \citenamefont {Zhang},\ and\ \citenamefont {Wu}}]{Lee.Zhang.Wu:09}%
  \BibitemOpen
  \bibfield  {author} {\bibinfo {author} {\bibfnamefont {W.-C.}\ \bibnamefont
  {Lee}}, \bibinfo {author} {\bibfnamefont {S.-C.}\ \bibnamefont {Zhang}}, \
  and\ \bibinfo {author} {\bibfnamefont {C.}~\bibnamefont {Wu}},\ }\href
  {\doibase 10.1103/PhysRevLett.102.217002} {\bibfield  {journal} {\bibinfo
  {journal} {Phys. Rev. Lett.}\ }\textbf {\bibinfo {volume} {102}},\ \bibinfo
  {pages} {217002} (\bibinfo {year} {2009})}\BibitemShut {NoStop}%
\bibitem [{\citenamefont {Platt}\ \emph {et~al.}(2012)\citenamefont {Platt},
  \citenamefont {Thomale}, \citenamefont {Honerkamp}, \citenamefont {Zhang},\
  and\ \citenamefont {Hanke}}]{Zhang2}%
  \BibitemOpen
  \bibfield  {author} {\bibinfo {author} {\bibfnamefont {C.}~\bibnamefont
  {Platt}}, \bibinfo {author} {\bibfnamefont {R.}~\bibnamefont {Thomale}},
  \bibinfo {author} {\bibfnamefont {C.}~\bibnamefont {Honerkamp}}, \bibinfo
  {author} {\bibfnamefont {S.-C.}\ \bibnamefont {Zhang}}, \ and\ \bibinfo
  {author} {\bibfnamefont {W.}~\bibnamefont {Hanke}},\ }\href {\doibase
  10.1103/PhysRevB.85.180502} {\bibfield  {journal} {\bibinfo  {journal} {Phys.
  Rev. B}\ }\textbf {\bibinfo {volume} {85}},\ \bibinfo {pages} {180502}
  (\bibinfo {year} {2012})}\BibitemShut {NoStop}%
\bibitem [{\citenamefont {Thomale}\ \emph {et~al.}(2011)\citenamefont
  {Thomale}, \citenamefont {Platt}, \citenamefont {Hanke}, \citenamefont {Hu},\
  and\ \citenamefont {Bernevig}}]{Thomale}%
  \BibitemOpen
  \bibfield  {author} {\bibinfo {author} {\bibfnamefont {R.}~\bibnamefont
  {Thomale}}, \bibinfo {author} {\bibfnamefont {C.}~\bibnamefont {Platt}},
  \bibinfo {author} {\bibfnamefont {W.}~\bibnamefont {Hanke}}, \bibinfo
  {author} {\bibfnamefont {J.}~\bibnamefont {Hu}}, \ and\ \bibinfo {author}
  {\bibfnamefont {B.~A.}\ \bibnamefont {Bernevig}},\ }\href {\doibase
  10.1103/PhysRevLett.107.117001} {\bibfield  {journal} {\bibinfo  {journal}
  {Phys. Rev. Lett.}\ }\textbf {\bibinfo {volume} {107}},\ \bibinfo {pages}
  {117001} (\bibinfo {year} {2011})}\BibitemShut {NoStop}%
\bibitem [{\citenamefont {Maiti}\ and\ \citenamefont
  {Chubukov}(2013)}]{Chubukov2}%
  \BibitemOpen
  \bibfield  {author} {\bibinfo {author} {\bibfnamefont {S.}~\bibnamefont
  {Maiti}}\ and\ \bibinfo {author} {\bibfnamefont {A.~V.}\ \bibnamefont
  {Chubukov}},\ }\href {\doibase 10.1103/PhysRevB.87.144511} {\bibfield
  {journal} {\bibinfo  {journal} {Phys. Rev. B}\ }\textbf {\bibinfo {volume}
  {87}},\ \bibinfo {pages} {144511} (\bibinfo {year} {2013})}\BibitemShut
  {NoStop}%
\bibitem [{\citenamefont {Carlstr\"om}\ \emph {et~al.}(2011)\citenamefont
  {Carlstr\"om}, \citenamefont {Garaud},\ and\ \citenamefont {Babaev}}]{Johan}%
  \BibitemOpen
  \bibfield  {author} {\bibinfo {author} {\bibfnamefont {J.}~\bibnamefont
  {Carlstr\"om}}, \bibinfo {author} {\bibfnamefont {J.}~\bibnamefont {Garaud}},
  \ and\ \bibinfo {author} {\bibfnamefont {E.}~\bibnamefont {Babaev}},\ }\href
  {\doibase 10.1103/PhysRevB.84.134518} {\bibfield  {journal} {\bibinfo
  {journal} {Phys. Rev. B}\ }\textbf {\bibinfo {volume} {84}},\ \bibinfo
  {pages} {134518} (\bibinfo {year} {2011})}\BibitemShut {NoStop}%
\bibitem [{\citenamefont {Watanabe}\ \emph {et~al.}(2014)\citenamefont
  {Watanabe}, \citenamefont {Yamashita}, \citenamefont {Kawamoto},
  \citenamefont {Kurata}, \citenamefont {Mizukami}, \citenamefont {Ohta},
  \citenamefont {Kasahara}, \citenamefont {Yamashita}, \citenamefont {Saito},
  \citenamefont {Fukazawa}, \citenamefont {Kohori}, \citenamefont {Ishida},
  \citenamefont {Kihou}, \citenamefont {Lee}, \citenamefont {Iyo},
  \citenamefont {Eisaki}, \citenamefont {Vorontsov}, \citenamefont
  {Shibauchi},\ and\ \citenamefont {Matsuda}}]{SWaveHoleDoped1}%
  \BibitemOpen
  \bibfield  {author} {\bibinfo {author} {\bibfnamefont {D.}~\bibnamefont
  {Watanabe}}, \bibinfo {author} {\bibfnamefont {T.}~\bibnamefont {Yamashita}},
  \bibinfo {author} {\bibfnamefont {Y.}~\bibnamefont {Kawamoto}}, \bibinfo
  {author} {\bibfnamefont {S.}~\bibnamefont {Kurata}}, \bibinfo {author}
  {\bibfnamefont {Y.}~\bibnamefont {Mizukami}}, \bibinfo {author}
  {\bibfnamefont {T.}~\bibnamefont {Ohta}}, \bibinfo {author} {\bibfnamefont
  {S.}~\bibnamefont {Kasahara}}, \bibinfo {author} {\bibfnamefont
  {M.}~\bibnamefont {Yamashita}}, \bibinfo {author} {\bibfnamefont
  {T.}~\bibnamefont {Saito}}, \bibinfo {author} {\bibfnamefont
  {H.}~\bibnamefont {Fukazawa}}, \bibinfo {author} {\bibfnamefont
  {Y.}~\bibnamefont {Kohori}}, \bibinfo {author} {\bibfnamefont
  {S.}~\bibnamefont {Ishida}}, \bibinfo {author} {\bibfnamefont
  {K.}~\bibnamefont {Kihou}}, \bibinfo {author} {\bibfnamefont {C.~H.}\
  \bibnamefont {Lee}}, \bibinfo {author} {\bibfnamefont {A.}~\bibnamefont
  {Iyo}}, \bibinfo {author} {\bibfnamefont {H.}~\bibnamefont {Eisaki}},
  \bibinfo {author} {\bibfnamefont {A.~B.}\ \bibnamefont {Vorontsov}}, \bibinfo
  {author} {\bibfnamefont {T.}~\bibnamefont {Shibauchi}}, \ and\ \bibinfo
  {author} {\bibfnamefont {Y.}~\bibnamefont {Matsuda}},\ }\href {\doibase
  10.1103/PhysRevB.89.115112} {\bibfield  {journal} {\bibinfo  {journal} {Phys.
  Rev. B}\ }\textbf {\bibinfo {volume} {89}},\ \bibinfo {pages} {115112}
  (\bibinfo {year} {2014})}\BibitemShut {NoStop}%
\bibitem [{\citenamefont {Okazaki}\ \emph {et~al.}(2012)\citenamefont
  {Okazaki}, \citenamefont {Ota}, \citenamefont {Kotani}, \citenamefont
  {Malaeb}, \citenamefont {Ishida}, \citenamefont {Shimojima}, \citenamefont
  {Kiss}, \citenamefont {Watanabe}, \citenamefont {Chen}, \citenamefont
  {Kihou}, \citenamefont {Lee}, \citenamefont {Iyo}, \citenamefont {Eisaki},
  \citenamefont {Saito}, \citenamefont {Fukazawa}, \citenamefont {Kohori},
  \citenamefont {Hashimoto}, \citenamefont {Shibauchi}, \citenamefont
  {Matsuda}, \citenamefont {Ikeda}, \citenamefont {Miyahara}, \citenamefont
  {Arita}, \citenamefont {Chainani},\ and\ \citenamefont
  {Shin}}]{SWaveHoleDoped2}%
  \BibitemOpen
  \bibfield  {author} {\bibinfo {author} {\bibfnamefont {K.}~\bibnamefont
  {Okazaki}}, \bibinfo {author} {\bibfnamefont {Y.}~\bibnamefont {Ota}},
  \bibinfo {author} {\bibfnamefont {Y.}~\bibnamefont {Kotani}}, \bibinfo
  {author} {\bibfnamefont {W.}~\bibnamefont {Malaeb}}, \bibinfo {author}
  {\bibfnamefont {Y.}~\bibnamefont {Ishida}}, \bibinfo {author} {\bibfnamefont
  {T.}~\bibnamefont {Shimojima}}, \bibinfo {author} {\bibfnamefont
  {T.}~\bibnamefont {Kiss}}, \bibinfo {author} {\bibfnamefont {S.}~\bibnamefont
  {Watanabe}}, \bibinfo {author} {\bibfnamefont {C.-T.}\ \bibnamefont {Chen}},
  \bibinfo {author} {\bibfnamefont {K.}~\bibnamefont {Kihou}}, \bibinfo
  {author} {\bibfnamefont {C.~H.}\ \bibnamefont {Lee}}, \bibinfo {author}
  {\bibfnamefont {A.}~\bibnamefont {Iyo}}, \bibinfo {author} {\bibfnamefont
  {H.}~\bibnamefont {Eisaki}}, \bibinfo {author} {\bibfnamefont
  {T.}~\bibnamefont {Saito}}, \bibinfo {author} {\bibfnamefont
  {H.}~\bibnamefont {Fukazawa}}, \bibinfo {author} {\bibfnamefont
  {Y.}~\bibnamefont {Kohori}}, \bibinfo {author} {\bibfnamefont
  {K.}~\bibnamefont {Hashimoto}}, \bibinfo {author} {\bibfnamefont
  {T.}~\bibnamefont {Shibauchi}}, \bibinfo {author} {\bibfnamefont
  {Y.}~\bibnamefont {Matsuda}}, \bibinfo {author} {\bibfnamefont
  {H.}~\bibnamefont {Ikeda}}, \bibinfo {author} {\bibfnamefont
  {H.}~\bibnamefont {Miyahara}}, \bibinfo {author} {\bibfnamefont
  {R.}~\bibnamefont {Arita}}, \bibinfo {author} {\bibfnamefont
  {A.}~\bibnamefont {Chainani}}, \ and\ \bibinfo {author} {\bibfnamefont
  {S.}~\bibnamefont {Shin}},\ }\href {\doibase 10.1126/science.1222793}
  {\bibfield  {journal} {\bibinfo  {journal} {Science}\ }\textbf {\bibinfo
  {volume} {337}},\ \bibinfo {pages} {1314} (\bibinfo {year}
  {2012})}\BibitemShut {NoStop}%
\bibitem [{\citenamefont {Grinenko}\ \emph {et~al.}(2017)\citenamefont
  {Grinenko}, \citenamefont {Materne}, \citenamefont {Sarkar}, \citenamefont
  {Luetkens}, \citenamefont {Kihou}, \citenamefont {Lee}, \citenamefont
  {Akhmadaliev}, \citenamefont {Efremov}, \citenamefont {Drechsler},\ and\
  \citenamefont {Klauss}}]{Grinenko2017}%
  \BibitemOpen
  \bibfield  {author} {\bibinfo {author} {\bibfnamefont {V.}~\bibnamefont
  {Grinenko}}, \bibinfo {author} {\bibfnamefont {P.}~\bibnamefont {Materne}},
  \bibinfo {author} {\bibfnamefont {R.}~\bibnamefont {Sarkar}}, \bibinfo
  {author} {\bibfnamefont {H.}~\bibnamefont {Luetkens}}, \bibinfo {author}
  {\bibfnamefont {K.}~\bibnamefont {Kihou}}, \bibinfo {author} {\bibfnamefont
  {C.~H.}\ \bibnamefont {Lee}}, \bibinfo {author} {\bibfnamefont
  {S.}~\bibnamefont {Akhmadaliev}}, \bibinfo {author} {\bibfnamefont {D.~V.}\
  \bibnamefont {Efremov}}, \bibinfo {author} {\bibfnamefont {S.-L.}\
  \bibnamefont {Drechsler}}, \ and\ \bibinfo {author} {\bibfnamefont {H.-H.}\
  \bibnamefont {Klauss}},\ }\href
  {https://link.aps.org/doi/10.1103/PhysRevB.95.214511} {\bibfield  {journal}
  {\bibinfo  {journal} {Phys. Rev. B}\ }\textbf {\bibinfo {volume} {95}},\
  \bibinfo {pages} {214511} (\bibinfo {year} {2017})}\BibitemShut {NoStop}%
\bibitem [{\citenamefont {Maiti}\ \emph {et~al.}(2015)\citenamefont {Maiti},
  \citenamefont {Sigrist},\ and\ \citenamefont
  {Chubukov}}]{ChubukovMaitiSigrist}%
  \BibitemOpen
  \bibfield  {author} {\bibinfo {author} {\bibfnamefont {S.}~\bibnamefont
  {Maiti}}, \bibinfo {author} {\bibfnamefont {M.}~\bibnamefont {Sigrist}}, \
  and\ \bibinfo {author} {\bibfnamefont {A.}~\bibnamefont {Chubukov}},\ }\href
  {\doibase 10.1103/PhysRevB.91.161102} {\bibfield  {journal} {\bibinfo
  {journal} {Phys. Rev. B}\ }\textbf {\bibinfo {volume} {91}},\ \bibinfo
  {pages} {161102} (\bibinfo {year} {2015})}\BibitemShut {NoStop}%
\bibitem [{\citenamefont {Silaev}\ \emph {et~al.}(2015)\citenamefont {Silaev},
  \citenamefont {Garaud},\ and\ \citenamefont {Babaev}}]{Silaev.Garaud.ea:15}%
  \BibitemOpen
  \bibfield  {author} {\bibinfo {author} {\bibfnamefont {M.}~\bibnamefont
  {Silaev}}, \bibinfo {author} {\bibfnamefont {J.}~\bibnamefont {Garaud}}, \
  and\ \bibinfo {author} {\bibfnamefont {E.}~\bibnamefont {Babaev}},\ }\href
  {\doibase 10.1103/PhysRevB.92.174510} {\bibfield  {journal} {\bibinfo
  {journal} {Phys. Rev. B}\ }\textbf {\bibinfo {volume} {92}},\ \bibinfo
  {pages} {174510} (\bibinfo {year} {2015})}\BibitemShut {NoStop}%
\bibitem [{\citenamefont {Garaud}\ \emph {et~al.}(2016)\citenamefont {Garaud},
  \citenamefont {Silaev},\ and\ \citenamefont {Babaev}}]{Garaud.Silaev.ea:15}%
  \BibitemOpen
  \bibfield  {author} {\bibinfo {author} {\bibfnamefont {J.}~\bibnamefont
  {Garaud}}, \bibinfo {author} {\bibfnamefont {M.}~\bibnamefont {Silaev}}, \
  and\ \bibinfo {author} {\bibfnamefont {E.}~\bibnamefont {Babaev}},\ }\href
  {\doibase 10.1103/PhysRevLett.116.097002} {\bibfield  {journal} {\bibinfo
  {journal} {Phys. Rev. Lett.}\ }\textbf {\bibinfo {volume} {116}},\ \bibinfo
  {pages} {097002} (\bibinfo {year} {2016})}\BibitemShut {NoStop}%
\bibitem [{\citenamefont {Garaud}\ and\ \citenamefont
  {Babaev}(2014)}]{Garaud.Babaev:14}%
  \BibitemOpen
  \bibfield  {author} {\bibinfo {author} {\bibfnamefont {J.}~\bibnamefont
  {Garaud}}\ and\ \bibinfo {author} {\bibfnamefont {E.}~\bibnamefont
  {Babaev}},\ }\href {\doibase 10.1103/PhysRevLett.112.017003} {\bibfield
  {journal} {\bibinfo  {journal} {Phys. Rev. Lett.}\ }\textbf {\bibinfo
  {volume} {112}},\ \bibinfo {pages} {017003} (\bibinfo {year}
  {2014})}\BibitemShut {NoStop}%
\bibitem [{\citenamefont {Lin}\ \emph {et~al.}(2016)\citenamefont {Lin},
  \citenamefont {Maiti},\ and\ \citenamefont {Chubukov}}]{Lin2016}%
  \BibitemOpen
  \bibfield  {author} {\bibinfo {author} {\bibfnamefont {S.-Z.}\ \bibnamefont
  {Lin}}, \bibinfo {author} {\bibfnamefont {S.}~\bibnamefont {Maiti}}, \ and\
  \bibinfo {author} {\bibfnamefont {A.}~\bibnamefont {Chubukov}},\ }\href
  {https://link.aps.org/doi/10.1103/PhysRevB.94.064519} {\bibfield  {journal}
  {\bibinfo  {journal} {Phys. Rev. B}\ }\textbf {\bibinfo {volume} {94}},\
  \bibinfo {pages} {064519} (\bibinfo {year} {2016})}\BibitemShut {NoStop}%
\bibitem [{\citenamefont {Sonier}\ \emph {et~al.}(2000)\citenamefont {Sonier},
  \citenamefont {Brewer},\ and\ \citenamefont {Kiefl}}]{Sonier2000}%
  \BibitemOpen
  \bibfield  {author} {\bibinfo {author} {\bibfnamefont {J.~E.}\ \bibnamefont
  {Sonier}}, \bibinfo {author} {\bibfnamefont {J.~H.}\ \bibnamefont {Brewer}},
  \ and\ \bibinfo {author} {\bibfnamefont {R.~F.}\ \bibnamefont {Kiefl}},\
  }\href {https://link.aps.org/doi/10.1103/RevModPhys.72.769} {\bibfield
  {journal} {\bibinfo  {journal} {Rev. Mod. Phys.}\ }\textbf {\bibinfo {volume}
  {72}},\ \bibinfo {pages} {769} (\bibinfo {year} {2000})}\BibitemShut
  {NoStop}%
\bibitem [{\citenamefont {Mahyari}\ \emph {et~al.}(2014)\citenamefont
  {Mahyari}, \citenamefont {Cannell}, \citenamefont {Gomez}, \citenamefont
  {Tezok}, \citenamefont {Zelati}, \citenamefont {de~Mello}, \citenamefont
  {Yan}, \citenamefont {Mandrus},\ and\ \citenamefont
  {Sonier}}]{PhysRevB.89.020502}%
  \BibitemOpen
  \bibfield  {author} {\bibinfo {author} {\bibfnamefont {Z.~L.}\ \bibnamefont
  {Mahyari}}, \bibinfo {author} {\bibfnamefont {A.}~\bibnamefont {Cannell}},
  \bibinfo {author} {\bibfnamefont {C.}~\bibnamefont {Gomez}}, \bibinfo
  {author} {\bibfnamefont {S.}~\bibnamefont {Tezok}}, \bibinfo {author}
  {\bibfnamefont {A.}~\bibnamefont {Zelati}}, \bibinfo {author} {\bibfnamefont
  {E.~V.~L.}\ \bibnamefont {de~Mello}}, \bibinfo {author} {\bibfnamefont
  {J.-Q.}\ \bibnamefont {Yan}}, \bibinfo {author} {\bibfnamefont {D.~G.}\
  \bibnamefont {Mandrus}}, \ and\ \bibinfo {author} {\bibfnamefont {J.~E.}\
  \bibnamefont {Sonier}},\ }\href {\doibase 10.1103/PhysRevB.89.020502}
  {\bibfield  {journal} {\bibinfo  {journal} {Phys. Rev. B}\ }\textbf {\bibinfo
  {volume} {89}},\ \bibinfo {pages} {020502} (\bibinfo {year}
  {2014})}\BibitemShut {NoStop}%
\bibitem [{\citenamefont {Stanev}\ and\ \citenamefont {Te\ifmmode
  \check{s}\else \v{s}\fi{}anovi\ifmmode~\acute{c}\else
  \'{c}\fi{}}(2010)}]{StanevTesanovic}%
  \BibitemOpen
  \bibfield  {author} {\bibinfo {author} {\bibfnamefont {V.}~\bibnamefont
  {Stanev}}\ and\ \bibinfo {author} {\bibfnamefont {Z.}~\bibnamefont
  {Te\ifmmode \check{s}\else \v{s}\fi{}anovi\ifmmode~\acute{c}\else
  \'{c}\fi{}}},\ }\href {\doibase 10.1103/PhysRevB.81.134522} {\bibfield
  {journal} {\bibinfo  {journal} {Phys. Rev. B}\ }\textbf {\bibinfo {volume}
  {81}},\ \bibinfo {pages} {134522} (\bibinfo {year} {2010})}\BibitemShut
  {NoStop}%
\bibitem [{\citenamefont {Marciani}\ \emph {et~al.}(2013)\citenamefont
  {Marciani}, \citenamefont {Fanfarillo}, \citenamefont {Castellani},\ and\
  \citenamefont {Benfatto}}]{Benfatto}%
  \BibitemOpen
  \bibfield  {author} {\bibinfo {author} {\bibfnamefont {M.}~\bibnamefont
  {Marciani}}, \bibinfo {author} {\bibfnamefont {L.}~\bibnamefont
  {Fanfarillo}}, \bibinfo {author} {\bibfnamefont {C.}~\bibnamefont
  {Castellani}}, \ and\ \bibinfo {author} {\bibfnamefont {L.}~\bibnamefont
  {Benfatto}},\ }\href {\doibase 10.1103/PhysRevB.88.214508} {\bibfield
  {journal} {\bibinfo  {journal} {Phys. Rev. B}\ }\textbf {\bibinfo {volume}
  {88}},\ \bibinfo {pages} {214508} (\bibinfo {year} {2013})}\BibitemShut
  {NoStop}%
\bibitem [{\citenamefont {Garaud}\ \emph {et~al.}(2017)\citenamefont {Garaud},
  \citenamefont {Silaev},\ and\ \citenamefont {Babaev}}]{Garaud2017}%
  \BibitemOpen
  \bibfield  {author} {\bibinfo {author} {\bibfnamefont {J.}~\bibnamefont
  {Garaud}}, \bibinfo {author} {\bibfnamefont {M.}~\bibnamefont {Silaev}}, \
  and\ \bibinfo {author} {\bibfnamefont {E.}~\bibnamefont {Babaev}},\ }\href
  {http://www.sciencedirect.com/science/article/pii/S0921453416300983}
  {\bibfield  {journal} {\bibinfo  {journal} {Ninth international conference on
  Vortex Matter in nanostructured Superdonductors}\ }\textbf {\bibinfo {volume}
  {533}},\ \bibinfo {pages} {63} (\bibinfo {year} {2017})}\BibitemShut
  {NoStop}%
\bibitem [{\citenamefont {Saint-James}\ \emph {et~al.}(1970)\citenamefont
  {Saint-James}, \citenamefont {Thomas},\ and\ \citenamefont
  {Sarma}}]{Saint-James.Thomas.ea}%
  \BibitemOpen
  \bibfield  {author} {\bibinfo {author} {\bibfnamefont {D.}~\bibnamefont
  {Saint-James}}, \bibinfo {author} {\bibfnamefont {E.}~\bibnamefont {Thomas}},
  \ and\ \bibinfo {author} {\bibfnamefont {G.}~\bibnamefont {Sarma}},\ }\href
  {http://books.google.fr/books?id=VeFjnQEACAAJ} {\emph {\bibinfo {title}
  {{Type II Superconductivity}}}},\ International series of monographs in
  natural philosophy\ (\bibinfo  {publisher} {Pergamon},\ \bibinfo {year}
  {1970})\BibitemShut {NoStop}%
\bibitem [{\citenamefont {Yuan}\ \emph {et~al.}(2009)\citenamefont {Yuan},
  \citenamefont {Singleton}, \citenamefont {Balakirev}, \citenamefont {Baily},
  \citenamefont {Chen}, \citenamefont {Luo},\ and\ \citenamefont
  {Wang}}]{Yuan2009}%
  \BibitemOpen
  \bibfield  {author} {\bibinfo {author} {\bibfnamefont {H.~Q.}\ \bibnamefont
  {Yuan}}, \bibinfo {author} {\bibfnamefont {J.}~\bibnamefont {Singleton}},
  \bibinfo {author} {\bibfnamefont {F.~F.}\ \bibnamefont {Balakirev}}, \bibinfo
  {author} {\bibfnamefont {S.~A.}\ \bibnamefont {Baily}}, \bibinfo {author}
  {\bibfnamefont {G.~F.}\ \bibnamefont {Chen}}, \bibinfo {author}
  {\bibfnamefont {J.~L.}\ \bibnamefont {Luo}}, \ and\ \bibinfo {author}
  {\bibfnamefont {N.~L.}\ \bibnamefont {Wang}},\ }\href
  {http://dx.doi.org/10.1038/nature07676} {\bibfield  {journal} {\bibinfo
  {journal} {Nature}\ }\textbf {\bibinfo {volume} {457}},\ \bibinfo {pages}
  {565} (\bibinfo {year} {2009})}\BibitemShut {NoStop}%
\bibitem [{\citenamefont {Tafti}\ \emph {et~al.}(2014)\citenamefont {Tafti},
  \citenamefont {Clancy}, \citenamefont {Lapointe-Major}, \citenamefont
  {Collignon}, \citenamefont {Faucher}, \citenamefont {Sears}, \citenamefont
  {Juneau-Fecteau}, \citenamefont {Doiron-Leyraud}, \citenamefont {Wang},
  \citenamefont {Luo}, \citenamefont {Chen}, \citenamefont {Desgreniers},
  \citenamefont {Kim},\ and\ \citenamefont {Taillefer}}]{PhysRevB.89.134502}%
  \BibitemOpen
  \bibfield  {author} {\bibinfo {author} {\bibfnamefont {F.~F.}\ \bibnamefont
  {Tafti}}, \bibinfo {author} {\bibfnamefont {J.~P.}\ \bibnamefont {Clancy}},
  \bibinfo {author} {\bibfnamefont {M.}~\bibnamefont {Lapointe-Major}},
  \bibinfo {author} {\bibfnamefont {C.}~\bibnamefont {Collignon}}, \bibinfo
  {author} {\bibfnamefont {S.}~\bibnamefont {Faucher}}, \bibinfo {author}
  {\bibfnamefont {J.~A.}\ \bibnamefont {Sears}}, \bibinfo {author}
  {\bibfnamefont {A.}~\bibnamefont {Juneau-Fecteau}}, \bibinfo {author}
  {\bibfnamefont {N.}~\bibnamefont {Doiron-Leyraud}}, \bibinfo {author}
  {\bibfnamefont {A.~F.}\ \bibnamefont {Wang}}, \bibinfo {author}
  {\bibfnamefont {X.-G.}\ \bibnamefont {Luo}}, \bibinfo {author} {\bibfnamefont
  {X.~H.}\ \bibnamefont {Chen}}, \bibinfo {author} {\bibfnamefont
  {S.}~\bibnamefont {Desgreniers}}, \bibinfo {author} {\bibfnamefont {Y.-J.}\
  \bibnamefont {Kim}}, \ and\ \bibinfo {author} {\bibfnamefont
  {L.}~\bibnamefont {Taillefer}},\ }\href {\doibase 10.1103/PhysRevB.89.134502}
  {\bibfield  {journal} {\bibinfo  {journal} {Phys. Rev. B}\ }\textbf {\bibinfo
  {volume} {89}},\ \bibinfo {pages} {134502} (\bibinfo {year}
  {2014})}\BibitemShut {NoStop}%
\bibitem [{\citenamefont {Landau}\ and\ \citenamefont
  {Lifshitz}(2013)}]{landau2013statistical}%
  \BibitemOpen
  \bibfield  {author} {\bibinfo {author} {\bibfnamefont {L.}~\bibnamefont
  {Landau}}\ and\ \bibinfo {author} {\bibfnamefont {E.}~\bibnamefont
  {Lifshitz}},\ }\href {https://books.google.fi/books?id=VzgJN-XPTRsC} {\emph
  {\bibinfo {title} {Statistical Physics}}},\ \bibinfo {number} {v. 5}\
  (\bibinfo  {publisher} {Elsevier Science},\ \bibinfo {year}
  {2013})\BibitemShut {NoStop}%
\end{thebibliography}%

%\end{thebibliography}%

\end{document}